\documentclass[a4paper,fleqn,usenatbib]{mnras}

\usepackage{txfonts}

\usepackage[T1]{fontenc}
\usepackage{ae,aecompl}

\usepackage{graphicx}	% Including figure files
\usepackage[usenames,dvipsnames]{xcolor}

\usepackage{savesym}
\savesymbol{iint}
\savesymbol{nnnfiiint}

\usepackage{wasysym}

\title[SN 1978K at three decades]{Evolving into a remnant: optical observations of SN 1978K at three decades}
\author[H. Kuncarayakti et al.]{H. Kuncarayakti$^{1,2}$\thanks{E-mail: hanin@das.uchile.cl},
K. Maeda$^{3,4}$,
J. P. Anderson$^{5}$,
M. Hamuy$^{2,1}$,
K. Nomoto$^{4,8}$,
\newauthor
L. Galbany$^{1,2}$,
M. Doi$^{6,7}$
\\
$^{1}$Millennium Institute of Astrophysics (MAS), Santiago, Chile \\
$^{2}$Departamento de Astronom\'ia, Universidad de Chile, Casilla 36-D, Santiago, Chile\\
$^{3}$Department of Astronomy, Kyoto University, Kitashirakawa-Oiwake-cho, Sakyo-ku, Kyoto 606-8502, Japan\\
$^{4}$Kavli Institute for the Physics and Mathematics of the Universe (WPI), University of Tokyo, Kashiwa, Chiba 277-8583, Japan\\    
$^{5}$European Southern Observatory, Alonso de C\'ordova 3107, Vitacura, Santiago, Chile \\
$^{6}$Institute of Astronomy, Graduate School of Science, The University of Tokyo, 2-21-1 Osawa, Mitaka, Tokyo 181-0015, Japan\\
$^{7}$Research Center for the Early Universe, Graduate School of Science, The University of Tokyo, Bunkyo-ku, Tokyo 113-003, Japan \\
}

\date{Accepted 2016 February 22. Received 2016 February 10; in original form 2015 November 04}

\pubyear{2016}

\begin{document}
\label{firstpage}
\pagerange{\pageref{firstpage}--\pageref{lastpage}}
\maketitle

\begin{abstract}
%[ D R A F T , t b w] \\
%\textcolor{violet}{} 
We present new optical observations of the supernova SN 1978K, obtained in 2007 and 2014 with the Very Large Telescope. We discover that the supernova has not faded significantly, {even} more than three decades after its explosion. The spectrum exhibits numerous narrow (FWHM $\lesssim600$ km s$^{-1}$) emission lines, indicating that the supernova blastwave is persistently interacting with dense circumstellar material (CSM). Evolution of emission lines indicates that the supernova ejecta is slowly progressing through the reverse shock, and has not expanded past the outer edge of the circumstellar envelope. We demonstrate that the CSM is not likely to be spherically distributed, {with mass of $\lesssim$ 1 M$_\odot$}. The progenitor mass loss rate {is} estimated as $\gtrsim 0.01$ M$_\odot$ yr$^{-1}$.
The slowly fading late-time light curve and spectra show striking similarity with SN 1987A, indicating that {a rate at which the CSM is being swept-up by the blastwave is gradually decaying and SN 1978K} is undergoing similar evolution to become a remnant. Due to its proximity (4~Mpc), SN 1978K {serves} as the next best example of late-time supernova evolution after SN 1987A.

%This is a simple template for authors to write new MNRAS papers.
%The abstract should briefly describe the aims, methods, and main results of the paper.
%It should be a single paragraph not more than 250 words (200 words for Letters).
%No references should appear in the abstract.
\end{abstract}

\begin{keywords}
supernovae: general -- supernovae: individual: SN 1978K% -- keyword3
\end{keywords}

%%%%%%%%%%%%%%%%%%%%%%%%%%%%%%%%%%%%%%%%%%%%%%%%%%

%%%%%%%%%%%%%%%%% BODY OF PAPER %%%%%%%%%%%%%%%%%%
\defcitealias{chugai95}{C95}
\defcitealias{ryder93}{R93}
\defcitealias{schlegel99}{S99}

\section{Introduction}
 \footnotetext[8]{Hamamatsu Professor}

SN 1978K was first discovered in 1990 as a strong H$\alpha$ source during a survey of H~\textsc{ii} regions in the nearby spiral galaxy NGC 1313 \citep{dopita90}. The object was initially classified as a nova, with a nebular spectrum exhibiting strong emission lines, until further observations suggested that it was actually a supernova (SN) well into the late-time phase \citep{ryder93}. The SN still radiates strongly in X-ray and radio wavelengths {for} many years after the supposed explosion in mid-1978, indicating {an on-going} interaction of the SN ejecta with dense circumstellar medium (CSM) \citep{ryder93,chugai95,schlegel99,gruendl02,chu99,lenz07,smith07}.

\citet{ryder93} reported that {a} possible progenitor of SN 1978K has been identified in pre-explosion photographic plates, having $B_J = 22.1$ mag, in 1974-1975. With their adopted distance to NGC 1313 of 4.5 Mpc ($\mu = 28.3$ {mag}) and assuming no reddening, that magnitude would correspond to a luminous blue object with absolute magnitude of $\sim -6$ {mag}. The object seemed to fade {below} $B_J \sim 23$ in Oct 1977, then came into outburst in 1978. Only four data points in the light curve were recovered from the 1978 outburst, with brightest magnitude $B = 16.0$ reached on 1978 Jul 31, and no spectroscopic data exist during the early phase. Therefore the actual SN type classification of this object is unknown although late-time observations show that the CSM is hydrogen-rich, suggesting a type-II event.

\citet{chu99} suggested that SN 1978K is associated with CSM similar to the ejecta nebulae of luminous blue variable (LBV) stars, based on the detection of the narrow components of H$\alpha$ and [N~\textsc{ii}] in high-resolution spectroscopy, which were interpreted as {having} originated in the unshocked CSM. This detection was also later confirmed by \citet{gruendl02}.
LBV stars such as $\eta$ Car and P Cyg are evolved massive stars that have been known to undergo irregular variability and eruptive mass loss episodes. A good number of SNe IIn have been associated with LBV progenitors \citep[see e.g. ][]{galyam09,smith11,kiewe12,taddia13}. 
While the connection between LBV stars and type-IIn SNe has often been suggested in the recent literature, alternatively red supergiant stars have also been proposed as one viable progenitor for this kind of objects \citep[e.g. SN 1998S,][]{mauerhan12}.

%IIn-LBV connection.
SN 1978K is arguably unique among other SNe observed at very late times. The emission lines are evidently narrow (full width at half maximum; FWHM $\sim$few hundred km s$^{-1}$) and long-lasting, with a wealth of various species in addition to the commonly detected H$\alpha$, H$\beta$, and oxygen lines. For comparison, SNe 1986J and 1979C observed at the age of $\sim$20 years only show a {few such} lines \citep{milisav08,milisav09}, which was also the case for SNe 1957D, 1970G, 1980K and 1993J \citep{milisav12}. Those SNe observed at late times typically show FWHM velocities in the order of several thousands km s$^{-1}$, significantly higher than that of SN 1978K. SN 1986J, for example, exhibits spectra containing dominant H$\alpha$ emission with {unchanging} width between 1986--1989 \citep[$\sim$4--7 years after the supposed explosion time,][]{leibundgut91}, and later observations in 1991 and 2007 show that the width had not changed significantly \citep{milisav08} although the line luminosity diminished greatly during that time period.

Optical spectra of SN 1978K were previously obtained in 1990 \citep[][henceforth R93]{ryder93}, 1992 \citep[][henceforth C95]{chugai95}, 1996 \citep[][henceforth S99]{schlegel99}, 1997 \citep{chu99}, and 2000 \citep{gruendl02}. The 1990--1996 spectra were obtained in low resolution ($\sim10$ \AA) while the 1997/2000 ones were in high resolution echelle spectroscopy ($\sim 0.3$ \AA). Here we report the spectroscopy of SN 1978K conducted in 2014 i.e. more than a decade after the last published spectrum and 36 years after the SN explosion, as well as archival spectra and photometry taken in 2007, and discuss the physical properties of the object derived from the observational data. 
The paper proceeds with Section \ref{sec:obs} describing the observations and data reduction, Section \ref{sec:result} with the results then Section \ref{sec:discussion} with the discussions, and ends with the Summary.

\section{Observations and data reduction}
\label{sec:obs}
SN 1978K was observed using the UT3/Melipal unit of the Very Large Telescope (VLT) at Cerro Paranal Observatory, Chile, and the VIMOS instrument in integral field unit (IFU) mode \citep{lefevre03}. The observation was done as part of an IFU survey\footnote{ESO observing program 094.D-0290 (PI: Kuncarayakti).} of nearby SN explosion sites to study the underlying stellar populations \citep{hk15}. Sky conditions during the observation on 2014 November 25 (UT) were photometric, with seeing varying between 0.5"--0.8" during the $2\times1800$s exposure of the object. The final image quality measured on the reduced datacube is around 1".

VIMOS was used in the IFU medium resolution mode with 13"$\times$13" field of view at the scale of 0.33" per spaxel. The spectral coverage is 4800--10000 \AA, with dispersion of 2.6~\AA/pixel. The effective line FWHM of the spectrum is $\approx$ 8~\AA. Spectrophotometric standard stars were also observed during the {same} night for the purpose of absolute flux calibration. The raw data were reduced using the Reflex\footnote{\url{https://www.eso.org/sci/software/reflex/}}-based VIMOS IFU pipeline \citep{freudling13}, resulting in two wavelength- and flux-calibrated datacubes in ($x,y,\lambda$) format for each of the 1800s exposures. The two datacubes were {averaged} and then the resulting final datacube {was} analysed using QFitsView\footnote{\url{http://www.mpe.mpg.de/~ott/dpuser/qfitsview.html}} \citep{ott12}. 

SN 1978K was identified in the IFU field of view and then the spectrum {was} extracted from the final science datacube using apertures with radius of 2 and 4 spaxels, corresponding to 0.66" and 1.32", respectively. These apertures were chosen as the radial profile of the SN is not perfectly Gaussian, with range in FWHM around 2--4 spaxels. 
{This is most likely an artifact of the observation and data reduction. The sky background was estimated and removed by using an annulus surrounding the extraction aperture.}
The spectrum resulting from 4-spaxels aperture radius is more noisy compared to that from 2-spaxels radius due to increased sky contamination. On the other hand, the 2-spaxels spectrum does not contain the whole flux from the object since the aperture misses the outer wings of the point-spread function. Therefore, we scaled the flux of the 2-spaxels spectrum to match the flux of the 4-spaxels one, with H$\alpha$ line flux as {a} reference. This scaled spectrum is used in the subsequent spectral analysis using {the \texttt{onedspec}} package in \textsc{Iraf}\footnote{\textsc{Iraf} is distributed by the National Optical Astronomy Observatory, which is operated by the Association of Universities for Research in Astronomy (AURA) under cooperative agreement with the National Science Foundation.}. Synthetic \textit{VRI} magnitudes were calculated from the spectrum, resulting in $V_{\textrm{synth}} = 20.7$, $R_{\textrm{synth}} = 19.3$, and $I_{\textrm{synth}} = 19.8$ with estimated errors of $\approx 0.2$ mag.

Additionally, we found and recovered raw photometric and spectroscopic data of SN 1978K in the ESO Science Archive Facility\footnote{\url{http://archive.eso.org/eso/eso_archive_main.html}}, taken from VLT/FORS2 \citep{appenzeller98} observations in 2007\footnote{ESO observing program 079.D-0124 (PI: Kjaer)}. The data were obtained in two different nights of observation: 2007 July 25 (photometry, \textit{BVRI} bands and spectroscopy, \textit{RIz} bands) and 2007 September 26 (spectroscopy, \textit{B} band) under clear sky conditions. 
Seeing varied between 0.9"--1.4" (photometry) and 0.6"--0.8" (spectroscopy). The spectroscopy uses the 1" longslit with the GRIS\_600B, GRIS\_600RI, and GRIS\_600z grism configurations. Exposure times were ${(2,4,4)} \times900$s for each {grism}, respectively.
This resulted in three spectra covering three wavelength regions (3500-6000 \AA, 5300-8500 \AA, 7500-10500 \AA) with dispersion of 0.8 \AA/pixel.
After raw data reduction using the Reflex-based FORS2 pipeline, both the spectroscopic and photometric data were flux calibrated using spectrophotometric and photometric standard star observations \citep{landolt09}. 
The three spectra were later combined together by averaging overlapping wavelength regions and analysed using \textsc{Iraf} as with the 2014 spectrum. Photometry was achieved using the \texttt{apphot} aperture photometry package within \textsc{Iraf}.
{The synthetic \textit{B} and \textit{I} magnitudes calculated from the spectrum agree with magnitudes resulting from direct imaging within 0.1 mag.}
The \textit{R}-band image however, was not usable as SN 1978K was saturated. Therefore, we use the synthetic \textit{R}-band magnitude for the subsequent analyses. The 2007 magnitudes are $B = 20.6$, $V = 20.4$, $R_{\textrm{synth}} = 19.2$, $I = 20.1$.

Throughout the paper we adopt the {distance to NGC 1313 as $4.61\pm0.21$ pc ($\mu = 28.3$ mag), according to the new Cepheid distance measurement using \textit{Hubble Space Telescope} by \citet{qing15}.}
This is slightly different compared to the value of 4.5 Mpc used in \citetalias{ryder93}, \citetalias{chugai95}, and \citetalias{schlegel99}, and 4.13 Mpc in \citet{smith07} and \citet{lenz07}.

\section{Results}
\label{sec:result}

\subsection{Light curve}

\begin{figure}
	\includegraphics[width=\linewidth]{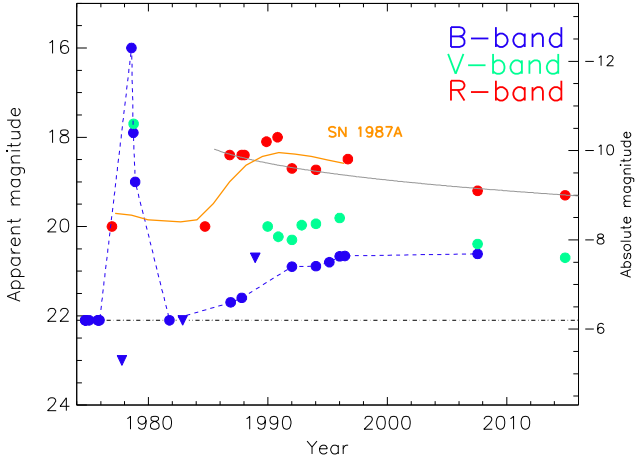}
    \caption{Light curve of SN 1978K between 1974-2014. The absolute magnitude axis assumes no reddening. The horizontal dash-dotted line indicates the apparent magnitude of the progenitor star. \textit{B}-band magnitudes are connected with {dashed} lines for clarity. Inverted triangles denote upper limits. Typical error in photometry is 0.3 mag. The solid orange line represents an approximation to SN 1987A \textit{R}-band light curve after the interaction with the CSM ring \citep{fransson15}, unscaled but shifted to match SN 1978K epoch and magnitudes. The {grey solid} line indicates power-law decay of the CSM interaction in SN 1978K (see text).}
    \label{lcurve}
\end{figure}

In Figure~\ref{lcurve} the historic light curve of SN 1978K is presented. Photometric data points from \citetalias{ryder93} and \citetalias{schlegel99} are plotted alongside our 2007/2014 photometry. As has been reported by \citetalias{ryder93}, the light curve evolution during the actual outburst event in 1978 was not well covered. Therefore, the peak in the light curve does not correspond to the actual event of maximum brightness of SN 1978K. 
Shortly prior to the 1978 eruption, the progenitor curiously became fainter in blue. While no information on the redder bands exist, it is possible that it expanded to become cooler and redder.

Looking at the {post-1978} light curve, it seems that there is a significant increase in brightness starting around mid-1980s. The $R$ magnitude brightened from 20.0 in September 1984, to 18.4 in October 1986. The $V$ and $B$-bands also seem to follow this behaviour. After the sudden increase, the luminosity went down gradually until our latest observation in 2014. This sudden increase is interpreted as the onset of the interaction between the SN blastwave and the surrounding CSM, analogous to what was experienced by SN 1987A. We discuss this interaction and comparison to SN 1987A in Section \ref{sec:discussion}, where the two SNe show analogous late time light curve{s} and spectra.

\subsection{Optical spectrum}
\label{sec:spec}
\begin{figure*}
	\includegraphics[width=\linewidth]{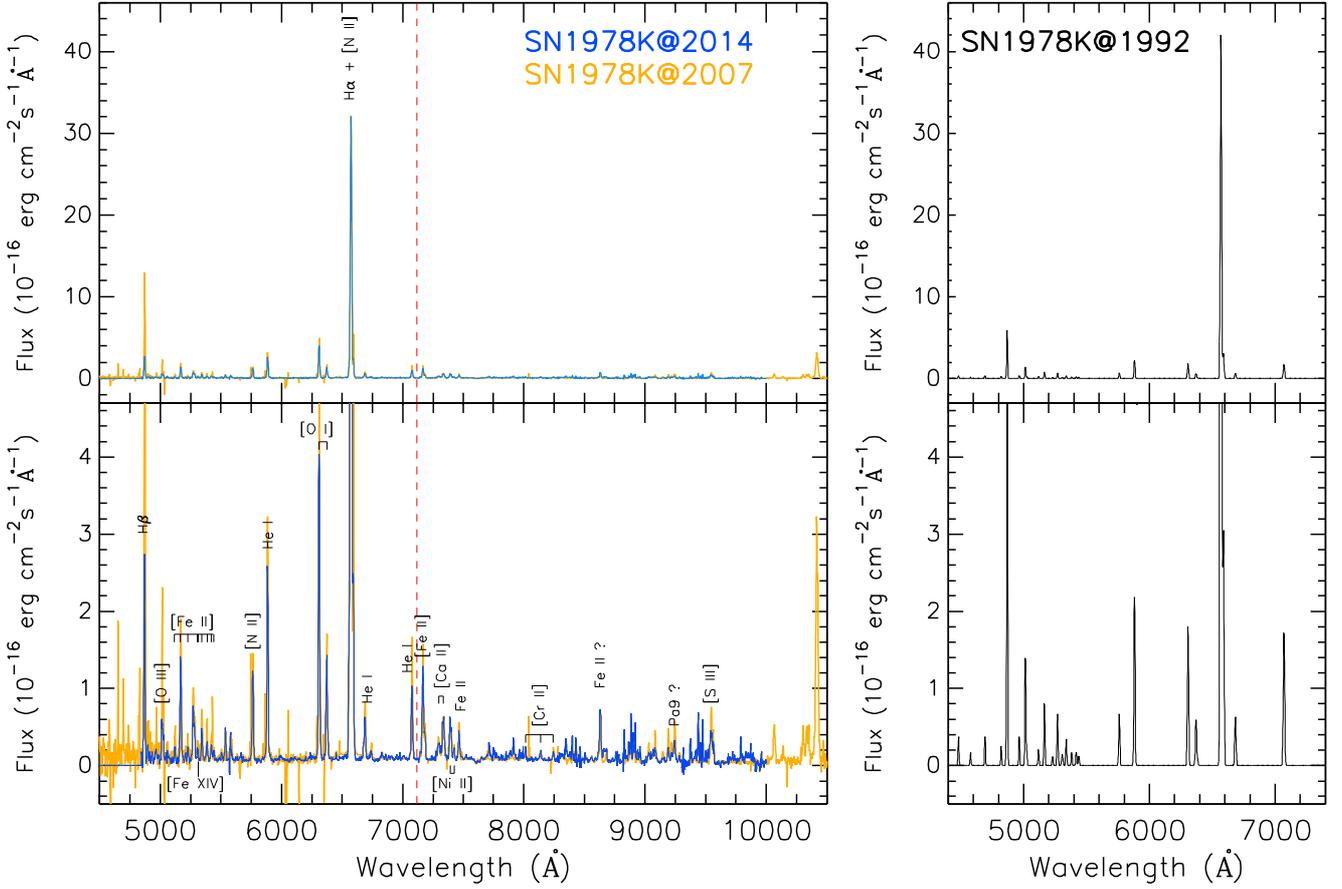}
    \caption{Observed spectra of SN 1978K in November 2014 and July 2007 (left panels), compared to the 1992 spectrum (right panels). Both left and right panels share the same scale in the plots. Upper panels shows the full coverage of the flux axis while bottom panels show the zoom-in to better examine the weaker lines. Spectra are not reddening-corrected. The red vertical dashed line indicates the red wavelength cutoff of the 1992 spectrum.}
    \label{spec78k}
\end{figure*}

The FORS2 and VIMOS spectra of SN 1978K at year 2007 and 2014 {are} shown in Figure~\ref{spec78k}, together with a synthetic spectrum representing {the one taken in 1992 \citepalias{chugai95}} for comparison. 
{The original 1992 EFOSC2 raw data were not available for retrieval in the ESO archives.}
{Therefore, the} 1992 spectrum was generated by constructing Gaussian curves with FWHM of 560 km s$^{-1}$, the observed H$\alpha$ FWHM which is comparable to the instrumental FWHM of the 1992 spectrum, and using tabulated wavelength and line fluxes in \citetalias{chugai95} to position the lines along the wavelength axis and scale the line strengths. 

Overall, the spectra show very similar appearances, which indicates that the same {physical process is still continuing} after two decades. The spectra are dominated by narrow emission lines,  in particular H$\alpha$ is the strongest, and also numerous other lines from various elements primarily helium, oxygen, and iron. The spectra of \citetalias{ryder93} and \citetalias{schlegel99} also show {a} similar appearance with the spectra displayed here in Figure \ref{spec78k}, although of a somewhat lower signal-to-noise ratio. 

We measure the line fluxes in the 2007/2014 spectra by fitting Gaussian profiles to the lines, after removing the very weak continuum component by fitting a linear function. We also applied deblending process in case two nearby lines are not well separated. Table~\ref{linetable} lists the lines identified in the spectrum. Line identifications follow that of SN 1987A \citep{groningsson08} and SN 2009kn \citep{kankare12}.
We identified most of the lines within our spectral wavelength range originally detected in 1992, with the exception of the ambiguous [Fe \textsc{x}] coronal line. [Fe \textsc{x}] $\lambda$6374 is possibly blended with the much stronger [O \textsc{i}] $\lambda$6364 line. Another coronal line, [Fe \textsc{xiv}] $\lambda$5303 is clearly detected, although blended with the nearby [Fe \textsc{ii}] $\lambda$5297. However, the two lines are of  comparable strengths, and therefore could be more easily deblended compared to the [Fe \textsc{x}]--[O \textsc{i}] blending. These two coronal lines are indicative of a hot ($\sim10^6$ K) shocked gas \citepalias{chugai95}.

\begin{table}
	\centering
	\caption{{Observed line strengths relative to $F(H\alpha) = 1000$.}}
	\label{linetable}
	\begin{tabular}{lcrr} % four columns, alignment for each
		\hline
%		A & B & C & D\\
Ion & $\lambda_0$ [\AA] & $F/F(H\alpha)$ & $F/F(H\alpha)$ \\
 & & 2007 & 2014 \\
		\hline
%		H9 & 3835.4 & & --- \\
%		H8 & 3889.1 &  & --- \\
%		H$\epsilon$ & 3970.1 &  & --- \\
%		H$\delta$ & 4101.7 &  & --- \\
%		H$\gamma$ & 4340.5 &  & --- \\
		H$\beta$ & 4861.3 & 250.2 & 67.3 \\
		$[$O \textsc{iii}] & 4958.9 & 7.4 & 3.8 \\
		$[$O \textsc{iii}] & 5006.8 & 27.2 & 11.3 \\
		He \textsc{i} & 5015.7 &	9.3 & 6.7 \\
		$[$Fe \textsc{ii}] & 	5111.6 &	5.2 & 3.7 \\
		$[$Fe \textsc{ii}] &	5158.8	& 33.8 & 36.0 \\
		$[$Fe \textsc{ii}] &	5220.1 &	2.1 & 3.4 \\
		$[$Fe \textsc{ii}] & 	5261.6 & 	27.2 & 27.1 \\
%		$[$Fe \textsc{ii}] &	5296.8 &	5.3 & 1.6 \\
		$[$Fe \textsc{xiv}] & 5302.9 & 5.3 & 1.5 \\
		$[$Fe \textsc{ii}] &	5333.6 &	11.6 & 9.4 \\
		$[$Fe \textsc{ii}] &	5376.5 &	11.1 & 6.3 \\
		$[$Fe \textsc{ii}] &	5413.0 &	15.7 & 4.7 \\
		$[$Fe \textsc{ii}] &	5433.1	& 2.9 &	2.8 \\
		$[$Fe \textsc{ii}] &	5527.6	& 6.6	&	9.4 \\
		$[$N \textsc{ii}] & 5754.5 &	30.6 & 33.6 \\
		He \textsc{i} & 5875.7 & 72.6 & 70.5 \\
		$[$O \textsc{i}] & 6300.3 & 101.8 & 113.0 \\
		$[$O \textsc{i}] & 6363.8 & 32.7 & 44.9 \\
%		$[$Fe X] &	6374.5	& --- & --- \\
		H$\alpha$ & 6562.8 & 1000.0 & 1000.0 \\
		$[$N \textsc{ii}] & 6583.3 &	80.3 & 65.7 \\
		He \textsc{i} & 6678.2 &	19.9 & 20.3 \\
		$[$S \textsc{ii}] & 6716.4 & 1.1 &	0.7 \\
		$[$S \textsc{ii}] & 6730.8 & 3.3 & 3.6 \\
		He \textsc{i} & 7065.7 & 38.8 & 31.8 \\
		$[$Fe \textsc{i}] & 7155.2 & 38.2 &	37.3 \\
		$[$Fe \textsc{ii}] & 7172.0 & 12.3 &	10.6 \\
		$[$Ca \textsc{ii}] (+ He I) & 7291.5 & 9.4 & 12.9 \\
		$[$Ca \textsc{ii}] (+ [O \textsc{ii}]) & 7323.9	& 24.9 & 25.9 \\
		$[$Ni \textsc{ii}] (+ [Fe \textsc{ii}]) & 7377.8	& 22.7 & 23.8 \\
		$[$Ni \textsc{ii}] &	7411.6 & 4.8 & 4.6 \\
		$[$Fe \textsc{ii}] &	7452.5 & 11.9 & 12.2 \\
%		K \textsc{i} &	7701.1 &	7714.0 & 6.2 \\
		$[$Cr~\textsc{ii}] & 8000.1 & 4.1 & 4.2 \\
		$[$Cr~\textsc{ii}] & 8125.3 & 3.0 & 3.4 \\
		$[$Cr~\textsc{ii}] & 8229.7 & 4.1 & 2.6 \\
		Fe~\textsc{ii}? & 8617 &	 16.6 & 24.7 \\
		?	& 9177 & 12.1 &	4.3 \\
		Pa9 + [Fe~\textsc{ii}] or Mg~\textsc{ii} & 9227 & 17.8 & 8.7 \\
		$[$S \textsc{iii}] & 9530.6 & 18.03 & 19.2 \\
		\hline
	\end{tabular}
\end{table}
%Observed H$\alpha$ flux is $4.4\times10^{-14}$ erg cm$^{-2}$ s$^{-1}$.

In the 2007/2014 spectra we also detected other lines redward of $\sim7000$ \AA, which were not detected or beyond the instrument response in the spectra taken in the 1990s. Redward of He \textsc{i} $\lambda$7065, we detected [Fe \textsc{ii}] $\lambda$7155, and also a complex of lines around 7200-7400 \AA. Among these lines, the [Ca \textsc{ii}] doublet of $\lambda\lambda$7291, 7324 {was} detected, with possible contamination from He \textsc{i} $\lambda$7281 and [O \textsc{ii}] $\lambda$ 7330. [Ni~\textsc{ii}] lines at  7378, 7412 \AA~are present. We also detect [Cr~\textsc{ii}] lines at 8000, 8125, 8230 \AA.
Further away we detected a strong line {at a rest wavelength of} 8617 \AA, possibly Fe~\textsc{ii} $\lambda$ 8620, another unidentified line at $\lambda_0 = 9177$ \AA, and a line at $\lambda_0 = 9227$ \AA. The latter could be a blend of Pa9 $\lambda$9229 and [Fe~\textsc{ii}]$\lambda9227$ \citep[see e.g.][]{groningsson08}, or Mg~\textsc{ii} $\lambda\lambda$9218, 9244  \citep{fransson13}. [S~\textsc{iii}] $\lambda$9531 is detected, very close to Pa8 $\lambda$9546.
In the 2007 spectrum we also detected Pa$\delta$ $\lambda$10049 and unidentified infrared lines around 10285, 10317, 10332 \AA, and a prominent line at 10399 \AA with strength comparable to He I $\lambda$5876.

Other lines of particular interest are {those} at observed wavelength $\sim$5537 \AA, and [O \textsc{i}] $\lambda$5577 whose presence was reported by \citetalias{ryder93} but not by \citetalias{chugai95} and \citetalias{schlegel99}. This [O \textsc{i}] $\lambda$5577 is most probably a night sky emission\footnote{\citetalias{ryder93} mentioned in their Table 3 that this line is blended with [O \textsc{i}] night sky line.} which was not clearly subtracted. This subtraction residual is also present in our spectra and the 1992 spectrum in \citetalias{chugai95}. Blueward of this line, the intriguing line at 5537 \AA~ was also clearly visible in the 1992 spectrum, but somehow not reported in the \citetalias{chugai95} paper. The line is not seen in the 1990 and 1996 spectra. Correcting from Doppler shift of 471 km s$^{-1}$ as measured from the H$\alpha$ line, the rest-frame wavelength of this line would be 5529 \AA.
{This line is probably the [Fe} \textsc{ii}{] $\lambda$5528 line as seen in SN 1987A spectrum and reported in \citet{groningsson08}}.

\subsection{Evolution of spectral line strengths}

\begin{figure}
	\includegraphics[width=\linewidth]{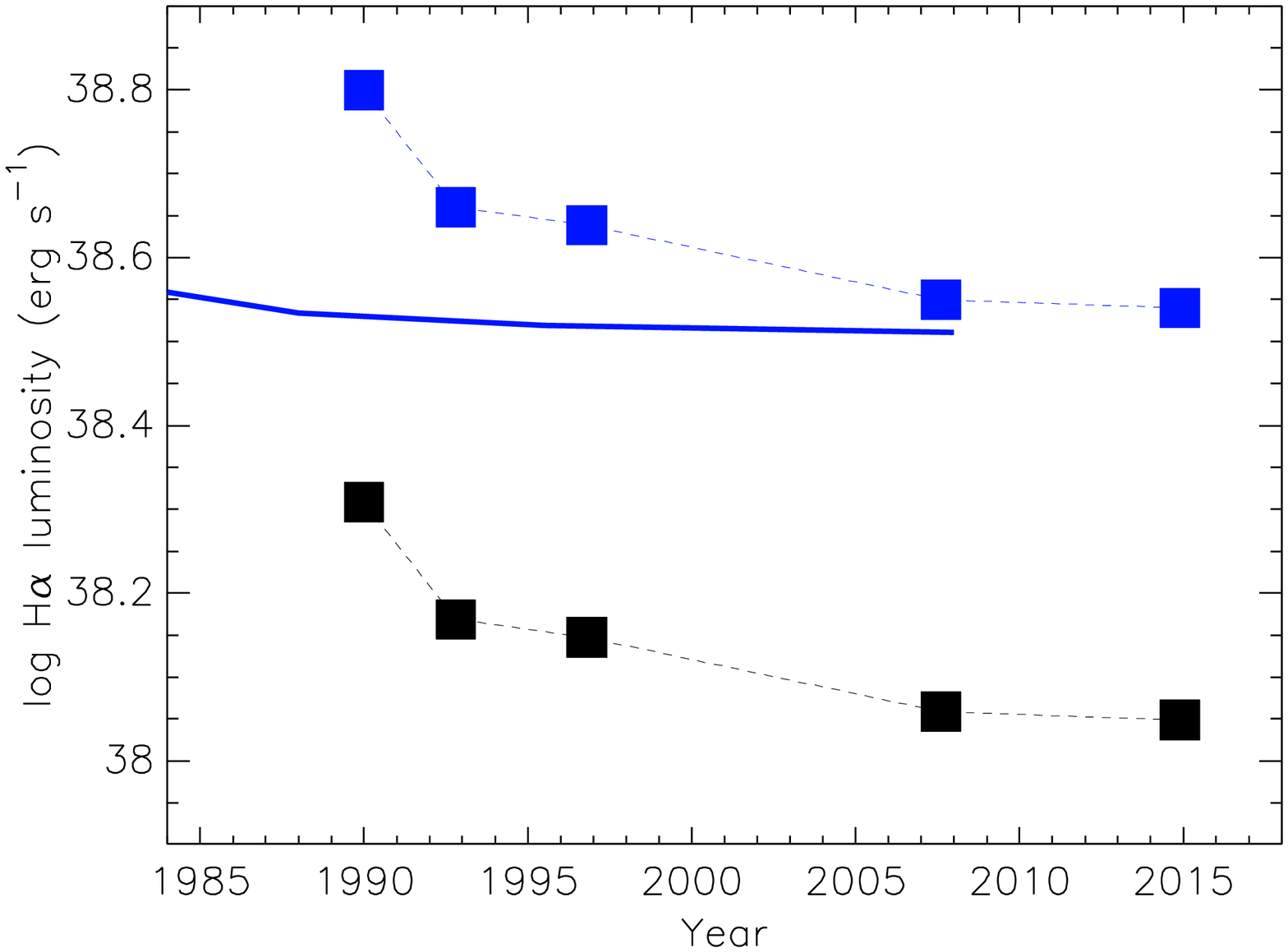}
    \caption{H$\alpha$ line luminosity evolution, 1990--2014. Black and blue data points show {observed} and dereddened flux, respectively. Blue solid curve denotes the prediction of H$\alpha$ luminosity evolution {from the Type II SN model of}  \citet{chevalier94}, shifted upward by two orders of magnitude.}
    \label{halum}
\end{figure}

We first examine the evolution of {the H$\alpha$ flux} as the strongest line in the optical regime. 
The observed H$\alpha$ flux in 2014 (2007) is $4.4$ $(4.5) \times 10^{-14}$ erg cm$^{-2}$ s$^{-1}$, corresponding to the {luminosity of $1.12$ $(1.14) \times 10^{38}$ erg s$^{-1}$.}
{As} with the 1992 spectrum where the line identifications are near-complete, H$\alpha$ contributes to roughly 60\% of the total line luminosity in the optical. Figure~\ref{halum} shows the evolution of H$\alpha$ luminosity from 1990 to 2014 {using also the spectra from \citetalias{ryder93}, \citetalias{chugai95}, and \citetalias{schlegel99}}. 
We dereddened the flux from a total line of sight absorption of $A_B = 2$ mag \citepalias{ryder93}, assuming $R_V = 3.1$ and interstellar extinction law of \citet{cardelli89}. We note that this corresponds to $A_V = 1.5$ mag, while \citetalias{chugai95} quoted $A_V = 0.64$ from the same \citetalias{ryder93} source. {The unreddened H$\alpha$ luminosity in 2014 (2007) is thus 3.47 (3.55) $\times 10^{38}$ erg s$^{-1}$.}

From Figure~\ref{halum} it is evident that the H$\alpha$ luminosity has been decreasing slowly since 1992. As reported by \citetalias{chugai95}, the H$\alpha$ luminosity peaked around 1990, from which point afterwards it decreases. Also in Figure~\ref{spec78k} it is evident that the H$\alpha$ line in 2014 is weaker compared to 1992, although not significantly. \citet{chevalier94} modeled late-time {emission} from type-II SNe interacting with circumstellar material. It was predicted that as the reverse shock weakens with time, the H$\alpha$ luminosity would also decrease steadily. We plot this prediction \citep[Table 6 of][]{chevalier94} in Figure~\ref{halum} as the blue solid line, after {scaling up by two} orders of magnitude and taking 1978.0 as the time of the explosion. The slowly decreasing behaviour of data points between 1990--2014 {resembles} that of the model, although the sharp drop from 1990 to 1992 does not.

\begin{figure}
	\includegraphics[width=\linewidth]{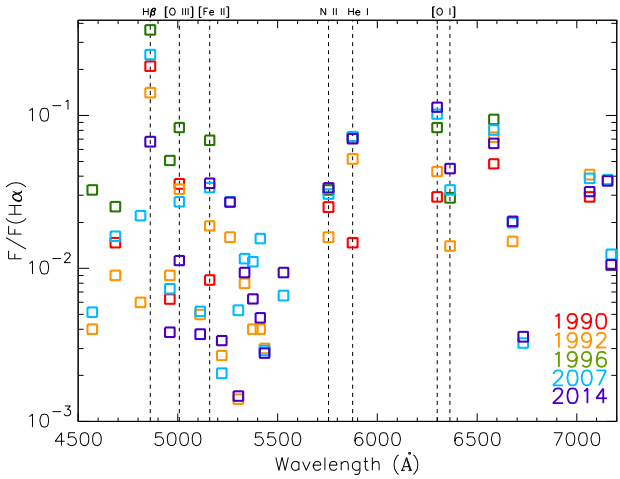}
    \caption{Evolution of line fluxes compared to H$\alpha$, from 1990--2014. Prominent lines are identified with vertical {dashed} lines. {The weak He} \textsc{i} {$\lambda$5016 line is not plotted for clarity, due to its proximity to [O}~\textsc{iii}{] $\lambda$5007.}
    }
    \label{linevol}
\end{figure}

To examine the evolution of other lines, we plot the line ratios with respect to H$\alpha$ from 1990--2014 in Figure~\ref{linevol}. In this plot, a systematic decrease in line ratio with time would result in a red-orange-green-blue-purple vertical sequence, and the opposite for an increase. From the plot, it is evident that the lines exhibit various behaviours.

The H$\beta$/H$\alpha$ ratio, or the inverse Balmer decrement, decreases from 1990 to 1992, but shoots up {in} 1996, decreases in 2007 and then drops again in 2014. The same behaviour is also seen in the [O~\textsc{iii}] line. It is to be noted that the spectra obtained in the 1990s were taken with slit spectroscopy that may have been affected by slit-loss and differential atmospheric refraction effects, which did not affect the 2014 spectrum taken with IFU spectroscopy. FORS2, while working in slit spectrosocpy mode, is equipped with an Atmospheric Dispersion Corrector. \citetalias{schlegel99} mentioned that their 1996 spectrum may have suffered from differential refraction, causing them to lose more of the red light compared to the blue. This may explain the behaviour seen in the ratio of lines bluer than 5500 \AA, where all the 1996 data points are higher up compared to the others. In the red part of the spectrum where the differential refraction effect is weaker the systematic behaviour is more visible. The [O \textsc{i}] doublet shows a systematic increase from 1990 to 2014. 
%The weak 1992 data point just redward of [O \textsc{i}] $\lambda$6364 is the purported [Fe \textsc{x}] line, which may have blended in the other spectra. 
Comparing the 1992 and {2007/}2014 spectra, it is apparent that generally the H$\alpha$ line becomes weaker relative to the helium, oxygen, and iron lines.

\subsection{SN 1978K environment}

SN 1978K exploded in the south-western outskirts of NGC 1313, a barred spiral galaxy. NGC 1313 is actively star-forming, and has been suspected to suffer interaction with a tidally disrupted satellite galaxy \citep[see e.g.][and references therein]{silva12}. 
{H} \textsc{i} {observations also show the presence of an expanding superbubble which is likely to have originated from this interaction \citep{ryder95}}.
The star formation rate is elevated in {the south-eastern} region of NGC 1313 and {the} star formation history suggests a recent, local starburst as opposed to a global starburst affecting the whole galaxy. SN 1978K lies within this south-western interaction region, thus it is quite likely that its progenitor was associated with the starburst. 

\begin{figure}
	\includegraphics[width=\linewidth]{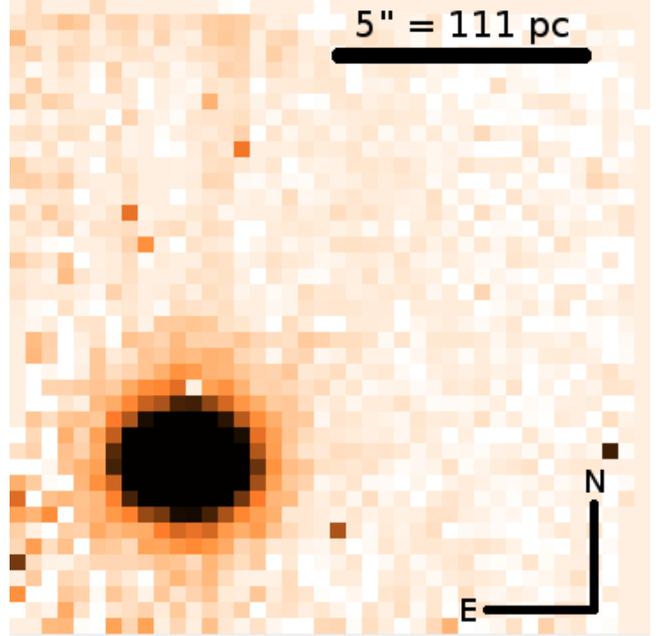}
    \caption{IFU field of view of SN 1978K in continuum-subtracted H$\alpha$ emission. The SN is the only visible source in the field.}
    \label{fovha}
\end{figure}

Our IFU data however, does not show any appreciable star formation {within $\sim50$ pc from SN 1978K and several hundred pc in the north and west directions (Figure~\ref{fovha})}. In any wavelength bin, whether in emission lines or continuum, the SN is the only source visible in the field. We do not detect any H$\alpha$ emission down to $\sim10^{-17}$ erg cm$^{-2}$ s$^{-1}$ \AA$^{-1}$, corresponding to H$\alpha$ luminosity of $\sim10^{35}$ erg s$^{-1}$. This implies that even a small H~\textsc{ii} region similar to the Orion nebula \citep[$L_{\textrm{H}\alpha} \approx 10^{37}$ erg s$^{-1}$,][]{crowther13} would have been well detected in our IFU data.

\section{Discussions}
\label{sec:discussion}
\subsection{Spectral evolution and the CSM}

\begin{figure}
	\includegraphics[width=\linewidth]{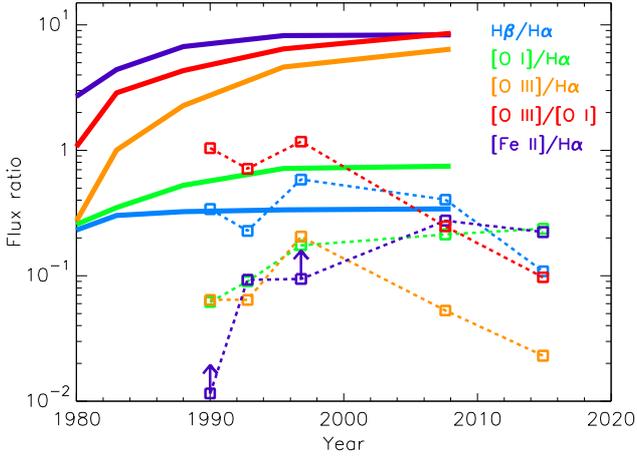}
    \caption{Line ratio evolution compared to the model by \citet{chevalier94}. The model is represented with solid lines, while the observed data points with squares. Arrows indicate lower limits for the [Fe \textsc{ii}] lines as the weaker lines were not detected in 1990 and 1996 spectra (see text for details).}
    \label{linevol2}
\end{figure}

\citetalias{chugai95} provided an explanation of the observed spectrum of SN 1978K as originating from radiative shock waves resulting from the interaction between the SN ejecta and dense clouds of circumstellar wind. In this picture the SN ejecta ploughs through the CSM, generating a reverse shock that is followed by a hot cooling region that is responsible for the H$\alpha$ emission and then the cool, partially ionized ejecta. This partially ionized region is responsible for the emission of low-ionization lines such as [Fe~\textsc{ii}] and [O~\textsc{i}] lines. As time passes, this outflowing, {more metal-rich region of SN ejecta} would overtake the reverse shock region and as a result the [Fe~\textsc{ii}] and [O~\textsc{i}] line luminosities would increase compared to H$\alpha$. This is exactly what is observed, as shown in Figure~\ref{linevol2}. 
{The model of emission from CSM interaction in type-II SNe by \citet{chevalier94} includes the evolution of line strengths.}
However, their model does not predict the same line ratio as observed (Figure~\ref{linevol2}). This discrepancy has also been reported by \citet{mauerhan12} who followed the evolution of SN 1998S at 14 years, and suggested that it probably stems from the uncertainty in the model to the shock velocity and ejecta density. Below we further discuss the discrepancy between the observed behaviours in SN 1978K and \cite{chevalier94}'s model.

Figure~\ref{linevol2} shows the evolution of line ratios H$\beta$/H$\alpha$, [O~\textsc{i}]/H$\alpha$, [O~\textsc{iii}]/H$\alpha$, [O~\textsc{iii}]/[O~\textsc{i}], and [Fe~\textsc{ii}]/H$\alpha$. Here the [O~\textsc{i}] flux is the combined flux of the doublet at $\lambda\lambda$6300, 6364, the [O~\textsc{iii}] combines $\lambda\lambda$4959, 5007, and [Fe~\textsc{ii}] combines all the flux of the [Fe~\textsc{ii}] lines. 
Upward arrows in the figure {denote} lower limits since in the 1990 and 1996 spectra only the [Fe~\textsc{ii}] $\lambda$5159 line was detected among the [Fe~\textsc{ii}] lines.
The ratio H$\beta$/H$\alpha$ is predicted to be increasing with time. The observed ratios closely match the predicted amount of the ratio although the behaviour is reversed, i.e. H$\beta$/H$\alpha$ is observed to be generally decreasing. 
For the other lines, the {amounts predicted} by the models deviate significantly from the observations. Also the line ratios are expected to be increasing with time, a behaviour that is only observed in [O~\textsc{i}]/H$\alpha$ and [Fe~\textsc{ii}]/H$\alpha$. 
The decreasing line ratios of the bluer lines with respect to H$\alpha$ could have been caused by dust formation, which cause them to suffer greater reddening compared to H$\alpha$. 
However, from the line profiles {alone there is no obvious evidence for} the presence of dust.

The model by \citet{chevalier94} assumes a CSM that was created by wind with smoothly decreasing density, {inversely} proportional to the radius squared, $r^{-2}$. In this model, the Balmer and the low-ionization lines are both associated to the cooling shell which is created by the interaction suffering a rapid cooling.
The high-ionization lines are emitted by the unshocked ejecta that are ionized by the high energy radiation from the interacting region. With a density decrease in the cooling shell, the Balmer decrement (H$\alpha$/H$\beta$) also decreases as H$\alpha$ becomes progressively optically thin. At the same time, it leads to the increasing importance of the forbidden lines.
In this case, similar behaviour would be displayed by \emph{both} the Balmer lines and the low-ionization (i.e. [O~\textsc{i}], [Fe~\textsc{ii}]) lines, either together they behave in accordance to the model prediction or opposite to it. 
The fact that the two lines show different behaviours suggests that there are different regions of {emission} for these lines and the CSM cannot be approximated with a simple spherically symmetric distribution. Furthermore, this is supported by the observed decrease of the inverse Balmer decrement, H$\beta$/H$\alpha$ ratio, which requires a density increase in the emitting region{, or else an increase in the reddening.}

\begin{figure}
	\includegraphics[width=\linewidth]{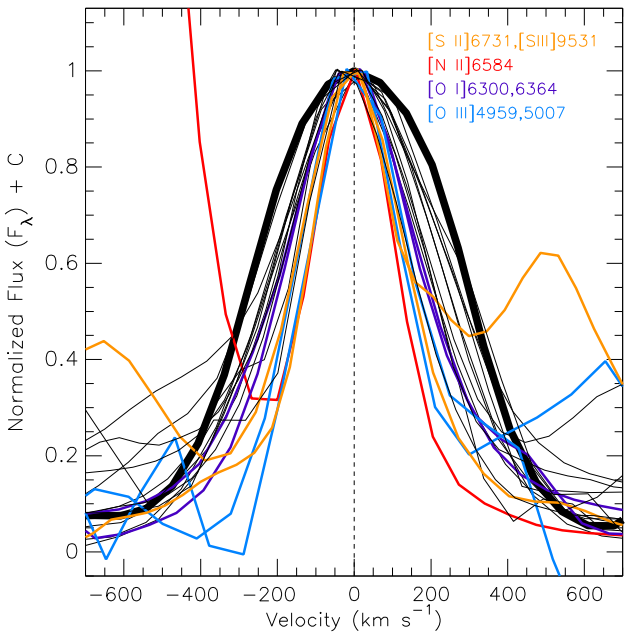}
    \caption{{Velocity} profiles of {the strongest} emission lines in the 2007 spectrum {that are not contaminated by a neighboring line}. H$\alpha$ is indicated with bold line {and the narrowest lines are colored.}}
    \label{lineprof}
\end{figure}

In Figure~\ref{lineprof} we plot the strongest lines in our 2007 spectrum in velocity space. It is clearly seen that most of the lines closely follow a symmetric, Gaussian-like profile with average FWHM of $500\pm100$ km s$^{-1}$. 
{The narrowest lines are} [O~\textsc{iii}], [S~\textsc{ii}], and [N~\textsc{ii}] forbidden lines. They exhibit velocity FWHM of 300--400 km s$^{-1}$, in contrast with the 580 km s$^{-1}$ velocity seen in H$\alpha$. These lines are commonly seen in normal H~\textsc{ii} regions, thus it is likely that they originate in the unshocked CSM further outside from the interaction region unlike what is assumed in the \citet{chevalier94} model. 

\citet{smith07} reported the {VLBI detection} of the remnant of SN 1978K. This remnant {was marginally resolved} and extends to $\sim10$ mas, corresponding to a diameter of 0.2 pc. If the ejecta {are} assumed to be expanding constantly at 500 km s$^{-1}$ since the time of the explosion as derived from the line FWHM, it would reach $\sim0.03$ pc size. However, the observed size is considerably larger thus it is more likely that the unimpeded SN ejecta expands with velocity in the order of several thousand km s$^{-1}$ as typically seen in other SNe. \citet{bartel07} showed that the outer boundary of radio emission is associated with the forward shock in a SN ejecta.
The $\sim500$ km s$^{-1}$ velocity seen in the emission lines corresponds to the expansion velocity of the emitting regions.
The narrow H$\alpha$ and [N \textsc{ii}] lines observed by \citet{chu99} using echelle spectroscopy show velocities in the order of 70-100 km s$^{-1}$, which is the velocity of the dense wind constituting the unshocked CSM. The data presented here do not have such high resolution to resolve the low-velocity components of the lines, therefore do not represent the true wind velocity of the unshocked CSM.

From the size of the remnant, \citet{smith07} estimated {an upper limit of} free-expansion velocity of $\sim4000$ km s$^{-1}$. If the expansion velocity remained constant, in the age of 36 years this remnant would further expand into $\sim0.3$ pc diameter, which is still much too small to be resolved with the IFU data at hand. 
{New \textit{VLBI} observations at 8.4 GHz has been obtained in 2015 by Ryder et al. (priv. comm.). Their data suggest that after 37 years the remnant is still barely resolved with a diameter of $<5$ mas ($<0.1$ pc), implying that the past average expansion velocity is actually $\lesssim 1500$ km s$^{-1}$.}
%> 0.5 * 0.005/206265*4.61e6*3.08567758e18/37/365/24/3600/1e5
%[1] 1477.602
{This is the upper limit of the past average expansion velocity and the actual current value would be smaller, though the exact value is unknown.}
{Since the expansion velocity of $\lesssim 1500$ km s$^{-1}$ is a 37-year average including both pre- and post-CSM crash velocities, this would imply a pre-crash free-expansion velocity of greater than 1500 km s$^{-1}$.}

{The picture of SN 1978K would be a SN ejecta freely expanding ($> 1500$ km s$^{-1}$)} and crashing into an inhomogeneous CSM of dense wind ($\sim100$ km s$^{-1}$) that was previously ejected by the progenitor star, and greatly decelerated resulting in shocks and emissions across the electromagnetic spectrum in the X-ray, optical, and radio wavelengths. 
As the CSM is inhomogeneously distributed, likely in clumps or in a ring or a combination of both, some part of the ejecta continues to travel unimpeded at {up to $\sim$ 1500 km s$^{-1}$ as high-velocity fragments, while the bulk of the ejecta expands at a somewhat lower velocity.}
The presence of low-velocity hydrogen and oxygen may indicate the mixing of hydrogen and oxygen from the high velocity to low velocity due to Rayleigh-Taylor instability. 

As time proceeds, the H$\alpha$ luminosity decreases due to the progressively smaller mass encountered by the ejecta. 
On the other hand, the H$\beta$/H$\alpha$ ratio decreases due to the shocked region {becoming} progressively of higher density, which argues against the $\rho \propto r^{-2}$ CSM and probably indicating a clumpy CSM.
The 500 km s$^{-1}$ velocity displayed by the optical emission lines represents the velocity of the decelerated forward shock or the reverse shock. Another possibility is that the CSM is approximately spherical, where {the $\lesssim 1500$ km s$^{-1}$ upper velocity limit derived from the radio represents the forward shock} and 500 km s$^{-1}$ represents the reverse shock. Nevertheless, in this situation typically the two velocities are similar, as long as the reverse shock is still in the outer steep density profile of the envelope. SN 1978K is well evolved, therefore it could be possible that the reverse shock already reached deep into the region in the ejecta where the density structure can be flat. In this case, it could be possible to have a large velocity difference between the forward and reverse shocks. However, as has been discussed above the evolution of the line ratios and the velocity profiles indicate that a spherically symmetric CSM is not plausible. 

The mass loss rate of the progenitor star can be estimated from the luminosity of the H$\alpha$ line, as the shock in the CSM is the dominant source of luminosity \citep{salamanca98}:

\begin{equation}
L_{H\alpha} = {1\over4} \epsilon_{H\alpha} {\dot{M} \over v_w} v_s^3 .
\end{equation}

Here if we adopt wind velocity $v_w$ = 70-100 km s$^{-1}$ and shock velocity $v_s$ = 500-600 km s$^{-1}$, the mass loss rate $\dot{M}$ would be in the order of 0.01 M$_\odot$ yr$^{-1}$. 
%> 2.9e38*4/.1*70e5/(500e5)^3*3600*24*365/2e33
This assumes {an} efficiency factor $\epsilon_{H\alpha} = 0.1$, which is actually applicable for young SNe and decreases to nearly zero with time. In the case of SN 1978K at the age of several decades, $\epsilon_{H\alpha}$ would be significantly smaller than 0.1, therefore the derived mass loss rate of $\sim0.01$ M$_\odot$ yr$^{-1}$ would be a lower limit. This implies that the progenitor of SN 1978K suffered heavy mass loss, with rate at least comparable to type-IIn SN progenitors with the highest mass loss rates \citep{kiewe12} and the averaged mass loss rate in the events of LBV giant eruptions \citep{smith14}. The comparison between the observed H$\alpha$ luminosity and the prediction from \citet{chevalier94} in Figure \ref{halum} show that the observed H$\alpha$ luminosity is about two orders of magnitude higher than the model. To first approximation, the model scales with $\dot{M}/v_w$, which assumes $\dot{M} = 50\times10^{-5}$ M$_\odot$ yr$^{-1}$ and $v_w = 100$ km s$^{-1}$. Scaling this by two orders of magnitude would require a mass loss rate in the order of $\sim$ 0.05 M$_\odot$ yr$^{-1}$ for similar wind velocities, which is quite consistent with the value calculated above.

The CSM density parameter $A_*$ is defined as proportional to $\dot{M} / v_w$, assuming a steady-state mass loss, where $A_* = 100$ for the case of $\dot{M} = 10^{-5}$ M$_\odot$ yr$^{-1}$ and $v_w = 10$ km s$^{-1}$ \citep{maeda12}. For SN 1978K, adopting the wind velocity and mass loss rate derived above, this would result in an extremely high density CSM with $A_* \sim 1.4 \times 10^4$. This is several orders of magnitude higher than the derived values for evolved massive stars such as Wolf-Rayet and red/yellow supergiant stars, and some well-studied type-IIb SN progenitors \citep{maeda15}.

Following equation (2) of \citet{patat95}, we estimate the amount of ionized hydrogen in the CSM of SN 1978K using the H$\alpha$ line luminosity. 
This equation essentially calculates the number of {recombinations} of ionized hydrogen that gives rise to the H$\alpha$ emission, and does not strongly depend on the CSM geometry.
Assuming model A velocity structure, the amount of ionized hydrogen at the age of 36 years would be $\sim0.7$ M$_\odot$ if electron {density is equal} to proton density ($n_e = n_{H^+}$). By comparing the relative flux $L_{H\alpha}/L_{He I} \propto 1.3n_{H^+}/n_{He^+}$ \citep[Case B recombination,][]{osterbrock06}, it is found that $n_{H^+}/n_{He^+} = 5.9$. This would mean that the CSM is composed of $\sim86 \%$ of H$+$ and $\sim14 \%$ of He$+$. Thus, the electron density $n_e$ would correspond to $100/86 \approx 1.2 n_{H^+}$. Adopting this value of electron density would increase the estimated H envelope mass by a factor of $\sqrt{1.2}$, {and} therefore would not significantly change the derived value of $M_{H^+} \sim 0.7$ M$_\odot$. 

Electron density can be estimated using line ratios from specific ions that were emitted by different levels with similar excitation energy, such as [S~\textsc{ii}] $\lambda6716 / \lambda6731$ \citep{osterbrock06}. The observed line ratio in SN 1978K is close to 0.2, which corresponds to high electron density exceeding $10^5$ cm$^{-3}$. \citet{chu99} estimated the electron density to be (3-12) $\times 10^5$ cm$^{-3}$. Using equation (4) of \citet{salamanca98}, the radius of the shock is estimated to be $\sim0.05$-0.10 pc if electron density is between 12 and 3 $\times 10^5$ cm$^{-3}$.
%> 1.54e16*(.1*90/(.01*70))^.5*(3e5/1e7)^(-.5) / 3.086e18
%[1] 0.1033086
%> 1.54e16*(.1*90/(.01*70))^.5*(12e5/1e7)^(-.5) / 3.086e18
%[1] 0.05165429
The explosion energy was then estimated by making use of equation (2.8) of \citet{chevalier94}. The equation calculates the shock radius from the mass loss rate, wind velocity, SN age, and reference density of the stellar density profile $\rho_r$. 
Assuming a flat outer density profile ($n = 7$), $\rho_r$ was estimated to be $0.05\times 10^{-16}$ g cm$^{-3}$ if the shock radius was 0.05 pc. A larger shock radius of 0.10 pc yields $\rho_r = 1.69 \times 10^{-16}$ g cm$^{-3}$.
Table 1 of \citet{chevalier94} gives the values of $\rho_r$ for set explosion energy of $E = 10^{51}$ erg and ejecta mass of $M = 10$ M$_\odot$. As the density scales with $E^2/M$ ($n=7$), the $\rho_r$ values are used to scale and estimate the explosion energy of SN 1978K to be in the range of $\sim$ 0.1-2{.0} $\times10^{51}$ erg, assuming an ejecta mass of 10 M$_\odot$.

Looking back at the historic light curve presented in Figure \ref{lcurve}, the sudden brightening in the {mid-1980s} is strikingly apparent.
If this sudden brightening is interpreted as the result of the interaction between the SN ejecta and the CSM, the CSM would be located at less than {0.01} pc away assuming an ejecta expanding at {1500} km s$^{-1}$. 
CSM located at this distance would require it to be produced by the dense progenitor wind {$\sim$150} years before the explosion. 
{As the average pre-shock expansion velocity is likely to be higher than 1500 km s$^{-1}$, the actual distance to the CSM would be larger thus it should have been formed earlier.}
{The interaction-powered luminosity evolution approximately follows a power-law decay,} represented with a dashed line in Figure \ref{lcurve}. Bumps above the line (data points around 1990 and 1996) may indicate that the CSM is clumpy, causing temporary brightness increases when the shock lights up the clumps.

{
%After the initial expansion up to interaction at $\sim$7 years post-SN, the ejecta is slowed down to 500 km s$^{-1}$. 
By assuming the volume of the CSM, the total mass of hydrogen in the CSM can be estimated by multiplying it with the hydrogen number density and the mass of a hydrogen atom. With a number density of $10^{5}$ cm$^{-3}$ {and assuming a spherically distributed CSM with an upper limit of radius of 0.05 pc as constrained by the 2015 \textit{VLBI} observations}, this will amount to about {1.3} M$_\odot$ of hydrogen in the CSM. However, as previously discussed the CSM geometry is likely to be non-spherical, and its size smaller and would not fully fill the sphere with 0.05 pc radius. Furthermore, the clumpy or ring-like CSM would not subtend a solid angle encompassing the entire surface of the shell therefore the actual CSM mass should not exceed {1.3} M$_\odot$.
Note that the amount of ionized hydrogen derived using the H$\alpha$ line luminosity is about 0.7 M$_\odot$, in agreement with this estimate.}
While \citetalias{ryder93} proposed a high CSM mass of $>80$ M$_\odot$ in a large ($\sim0.1$ pc) CSM shell, \citetalias{chugai95} proposed a clumpy model of CSM to derive the mass of $\sim1$ M$_\odot$. Our estimates are more consistent with a $\lesssim 1$ M$_\odot$ CSM, and concur with the picture proposed by \citet{chu99} where the SN ejecta is significantly decelerated by a dense, closer CSM although the exact geometry of the CSM is unknown.

%The total mass of hydrogen shell if interaction starts at 1985:
%interaction sphere vol: 4/3*pi*(4000e5*3600*24*365*7)^3 cm3
%plus shell with 500 km/s for next 29yr: 4/3*pi*( 4000e5*3600*24*365*7 + 500e5*3600*24*365*29)^3
%Shell volume = > 7.201065e+51
%Total H+ (=n_e) = shell vol*1e5
%Total H+ weight: > 7.2e51*1e5 * 1.67e-24 = 0.6 Msun

%with 1500 km/s:
%interaction sphere vol: 4/3*pi*(1500e5*3600*24*365*7)^3 cm3
%plus shell with 500 km/s for next 29yr: 4/3*pi*( 1500e5*3600*24*365*7 + 500e5*3600*24*365*29)^3
%shell vol = 1.900633e+51
%Total H+ (=n_e) = shell vol*1e5
%Total H+ weight: > 1.9e51*1e5 * 1.67e-24 = 0.2 Msun

%\subsection{Comparison with other SNe}
\subsection{Luminosities and comparison with other SNe}
The evolution of SN 1978K during the 1980s has been investigated by \citetalias{ryder93}. After the 1978 event, the object faded to its pre-SN magnitude and remained so from at least 1981, then slowly brightened again towards the end of the decade. The {843 MHz} radio light curve shows that it peaked around the year 1984 {\citepalias{ryder93}, while in the 5 GHz band the peak occurred in early 1981 \citepalias{schlegel99}}. 
Far-infrared \textit{IRAS} data shows that the SN was not detected in 1983, corresponding to an upper limit of $5 \times 10^{40}$ erg s$^{-1}$ at 10 $\mu$m. \citet{tanaka12} suggested the presence of shocked 
$1.3 \times 10^{-3}$ M$_\odot$ of circumstellar silicate dust in SN 1978K from the analysis of the infrared SED from 2006-2007 \textit{AKARI} and \textit{Spitzer} data. The derived infrared luminosity was $1.5 \times 10^{39}$ erg s$^{-1}$.

X-ray observations in 1980 with the \textit{Einstein} X-ray satellite did not detect the SN down to an unabsorbed luminosity $2.0\times10^{39}$ erg s$^{-1}$, while the 1991 \textit{ROSAT} observation showed that SN 1978K had brightened to $9.5 \times 10^{39}$ erg s$^{-1}$ in the 0.2-2.4 keV range. The most recent result by \citet{smith07} showed that the X-ray unabsorbed luminosity within 0.2-10 keV is $2.9 \times 10^{39}$ erg s$^{-1}$. 
\citet{dwarkadas14} showed that X-ray luminosities of core-collapse SNe are typically lower than $10^{39}$ erg s$^{-1}$ after $\sim10^4$ days, with SNe IIn exhibiting the highest luminosity compared to the other SN subclasses. This is comparable with the exceptionally high X-ray luminosity of SN 1978K, even decades after the explosion. 
Thermal X-ray emission flux increases with the square of wind density, thus the interpretation is consistent with SN 1978K {having} exploded into a circumstellar environment dominated by a dense wind typical {of} SNe IIn. The fact that SN 1978K did not show strong radio and X-ray emissions during the earliest years suggests that the immediate CSM could have been sparse\footnote{{The \textit{Einstein} satellite could have detected SN 1978K were it brighter, while at that time an interferometer like \textit{ATCA} was not yet available in order to resolve the SN flux from the host galaxy even if it was above the detection limit. The model light curve fits in Figure 10 of \citetalias{schlegel99} indicate a peak flux in the decade following the explosion of close to 1 Jy, which would have easily been detectable with e.g. the Parkes radio telescope were it well resolved.}}. 
This is probably analogous to SN 1996cr, which was suspected to explode in a cavity-like environment before eventually the ejecta strikes a dense surrounding CSM \citep{bauer08}. 
{The CSM around SNe nevertheless exhibit various structures. The type-II SN 1996al, for example, has recently been shown to interact with a complex CSM, consisting of a dense inner CSM and equatorial ring embedded in a less dense but clumpy halo \citep{benetti15}.}

\begin{figure}
	\includegraphics[width=\linewidth]{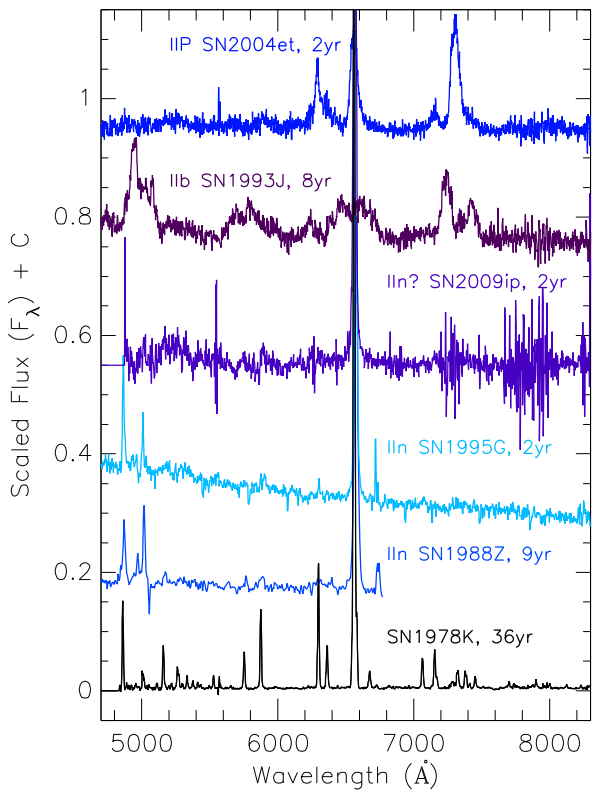}
    \caption{Comparison of 2014 SN 1978K spectrum with several type-II SNe observed at late times. The spectrum of SN 1995G is published in \citet{pastorello02}, SN 1988Z in \citet{aretxaga99}, SN 1993J in \citet{modjaz14}, and SN 2004et in \citet{sahu06}.}
    \label{spec2n}
\end{figure}

Figure~\ref{spec2n} shows the comparison of the late-time spectrum of SN 1978K to those of several type-II SNe. The spectrum of SN 2009ip\footnote{We note that the nature of SN 2009ip as a genuine SN explosion is still unclear {\citep[see e.g. ][]{fraser15,margutti14}}. The age of 2 years in Figure \ref{spec2n} is with reference to the 2012 event.} 
was obtained in 2014 in the same observing program as SN 1978K, using the identical VLT/VIMOS instrument setup. Spectra of SNe 1995G, 1988Z, and 2004et were obtained from the SUSPECT\footnote{\url{http://www.nhn.ou.edu/~suspect/}} Online Supernova Spectrum Archive, while SN 1993J was from the WISeREP\footnote{\url{http://wiserep.weizmann.ac.il/}} repository \citep{yaron12}. From the figure it is apparent that the {current} spectrum of SN 1978K {does} not resemble SNe IIb and IIP {during their first decade.}
It bears more similarities with some SNe IIn, although there are also notable differences. All of the spectra exhibit dominant Balmer lines, and also the indication of iron lines around $\sim$5200-5400 \AA. He \textsc{i} lines at $\lambda$5876 and $\lambda$7065 are also present in the SNe IIn. The dominant, narrow [N \textsc{ii}] $\lambda$5755, He \textsc{i} $\lambda$5876 and [O \textsc{i}] $\lambda\lambda$6300, 6364 are seen in SN 1978K.

The location of SN 1978K in the host galaxy is also somewhat consistent with type-IIn SNe on average. It is located in the host outskirts, nearly 6 kpc from the galaxy center and in the immediate surroundings of SN 1978K there is no star formation detected. 
This is similar {to} SN 2009ip which also exploded in the outskirts of its host galaxy, in a region with insignificant star formation. \citet{habergham14} showed that type-IIn SNe are weakly associated with ongoing star formation, a puzzling fact since these SNe are often associated with massive LBV stars. \citet{smith15} noticed the isolated nature of LBV stars in the Milky Way and Magellanic Clouds, and suggested that 	they are mass gainers that were kicked out of binary systems following the SN {explosion} of the mass donor companion.
Note that in general type-II SNe do not follow closely the ongoing star formation as traced by H~\textsc{ii} regions \citep{anderson12,galbany14}.

%09ip: LBV, no SF -- similar! spectra blm evolved enough. 09ip >4 kpc, 78k nearly 6 kpc from host center

\subsection{Comparison with SN 1987A}
\begin{figure*}
	\includegraphics[width=\linewidth]{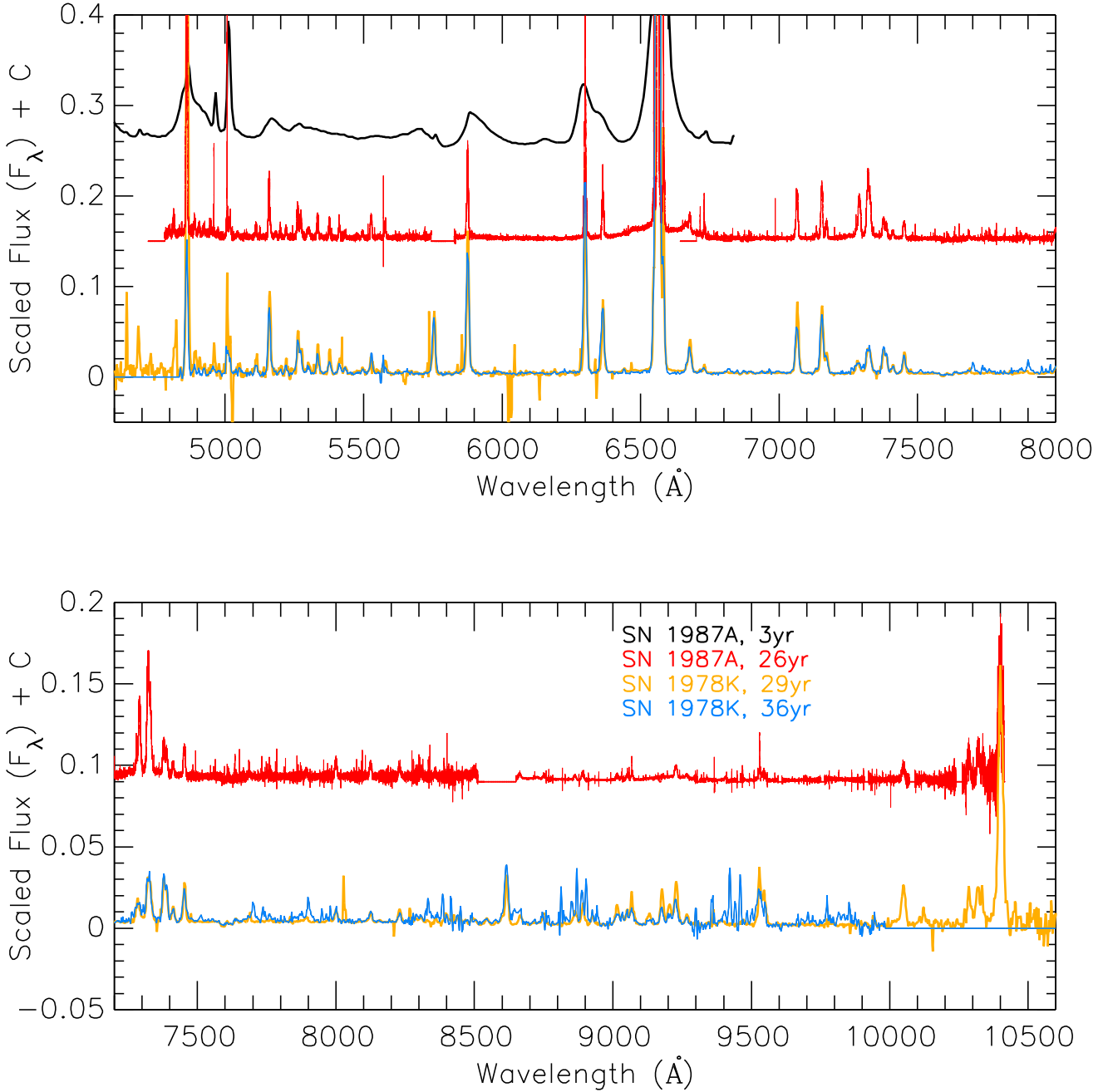}
    \caption{Comparison of SN 1978K spectrum with SN 1987A observed at late times. The nebular spectrum of SN 1987A was initially showing broad emission lines, which eventually becomes narrow similar to what is now observed on SN 1978K. The SN 1987A spectra includes the central ejecta and the CSM ring regions inside the slit \citep[see][]{fransson13}.}
    \label{spec87a}
\end{figure*}

It is interesting to compare the spectrum of SN 1978K with the nearest modern supernova, SN 1987A. Figure~\ref{spec87a} shows the comparison of SN 1978K at 2014 (age 36 years) with SN 1987A at 1990 and 2013 (age 3 and 26 years, respectively). We obtained the 1990 spectrum of SN 1987A \citep{pun95} from SUSPECT, while the 2013 VLT/UVES echelle spectrum was taken from the ESO Science Archive Facility, Phase 3 data release of observing program 092.D-0119 \citep{fransson13,fransson15}. SN 1987A, due to its proximity, is spatially resolved and has been very well monitored up until very late phase where the SN ejecta {interact} with the distinct ring-shaped CSM. The emergence of hot spots in the ring indicates the interaction between the blast wave (forward shock) and the CSM ring, which occurred first in 1995, or 8 years after the SN explosion \citep{lawrence00}. The interaction greatly decelerates the SN ejecta and generates reverse shock in the inner radius of the ring \citep{sonneborn98}. 

The comparison between SN 1978K and SN 1987A in Figure \ref{spec87a} immediately shows that the spectra of the two objects at the age of $\sim3$ decades look very similar. Nearly the same set of emission lines appear in both SNe, with similar relative strengths. This suggests that the SNe are in the same phase of evolution to become young SN remnants, and that similar physical processes are taking place. 
{In SN 1987A} we detect the [Fe \textsc{xiv}] $\lambda$5303 line and there is no indication of [Fe \textsc{x}] $\lambda$6374 {as is the case for SN 1978K (c.f. Section \ref{sec:spec})}.
As exemplified by SN 1987A where the initially broad spectral lines evolve to become much narrower, this indicates that indeed there is a presence of dense CSM around SN 1978K that is decelerating the expansion of the ejecta {($\sim$10$^3$-10$^4$ km s$^{-1}$ assuming typical SN ejecta velocities; initially $\sim$35 000 km s$^{-1}$ in the case of SN 1987A)} thus creating shocks and slowly declining post-interaction light curve (see Figure \ref{lcurve}). The striking similarity of the late-time light curves of SN 1978K and 1987A again suggests the similar processes of interaction in the two SNe although the time of the onsets may be different. The peak of SN 1987A interaction light curve occurred around day 8000 \citep{fransson15}, or 22 years after the SN, while in SN 1978K it occurred after $~\sim12$ years, indicating a relatively closer CSM concentration. This could further imply that the progenitor of SN 1978K changed {its} mass loss rate in a shorter timescale prior to the explosion compared to what SN 1987A progenitor did{, assuming similar ejecta velocities and progenitor wind speeds.}

In SN 1987A the shock propagates at $\sim$540 km s$^{-1}$ through the unshocked ($\sim$10 km s$^{-1}$) circumstellar ring and accelerates the post-shock gas \citep{fransson15}, resulting in typical line velocities of $\sim$300 km s$^{-1}$ \citep{groningsson08} in the shocked hot spots.
In this regard, one may even speculate that the CSM geometry in SN 1978K resembles the ring structure and the SN might be regarded as a more distant version of the SN 1987A remnant. \citetalias{ryder93} suggested that the peak magnitude of SN 1978K as a type-IIP or IIL SN could have reached $M_B \sim -14$ to $-15$ mag, with the caveat that the 1978 light curve is very poorly sampled, thus it could have been a subluminous or even a non-terminal explosion. Note that SN 1987A, now considered a peculiar type-II SN, is also somewhat underluminous with peak $M_V \approx -16$ mag \citep{schaeffer87}. 

The emission of radiation from the SN 1987A ring is now decaying as the interaction proceeds to destroy it \citep{fransson15}. The same process is possibly also happening in SN 1978K as the late-time light curve suggests. 
This corroborates the fact that the spectrum has changed very little {in the last} 20 years. Observations of this object in the following years/decades may reveal the weakening and eventual disappearance of the emissions from the CSM interaction. It would undoubtedly be interesting to resolve the remnant and CSM of SN 1978K and compare it to SN 1987A, although such observational effort would only be possible to be achieved using interferometry or the next generation 30m-class telescopes {operating at diffraction limit in the infrared}.

\section{Summary}
%The last numbered section should briefly summarise what has been done, and describe the final conclusions which the authors draw from their work.
We present late-time optical spectroscopy of SN 1978K in NGC 1313, obtained in 2007 and 2014\footnote{{Spectra available at the WISeREP database.}}. The spectrum still exhibits strong narrow emission lines (FWHM $\lesssim 600$ km s$^{-1}$) and has not changed much since the last published observations in the 1990s. We derive a progenitor mass loss rate of greater than 0.01 M$_\odot$ yr$^{-1}$ and CSM mass of less than {$\sim$ 1} M$_\odot$. Emission line ratios suggest that the CSM is more likely to be inhomogeneous, and the increasingly metal-rich inner ejecta is progressing to overtake the reverse shock.
The late-time light curve suggests that the interaction between SN blastwave and the CSM started around early to mid-1980s, and is currently decaying slowly. This behaviour is akin to SN 1987A, and the SN 1978K spectra are strikingly similar to that of SN 1987A at 26 years. We infer that SN 1978K is currently undergoing analogous process as SN 1987A in a similarly inhomogemeous circumstellar environment where CSM interaction decays as the SN proceeds to evolve into a remnant. Continuous monitoring of this interesting object in the coming years in all accessible wavelengths is strongly encouraged.

\section*{Acknowledgements}
We gratefully acknowledge the referee, Stuart Ryder, for thorough reading and useful suggestions that helped improve the paper significantly.
L. Dessart and S. Blinnikov are thanked for inspiring discussions.
Parts of this work were conducted while being hosted by Kavli IPMU, to which HK is grateful.
Support for HK, MH, and LG is provided by the Ministry of Economy, Development, and Tourism's Millennium Science Initiative through grant IC120009, awarded to The Millennium Institute of Astrophysics, MAS. HK and LG acknowledge support by CONICYT through FONDECYT grants 3140563 and 3140566. This work is based on ESO observing programs 094.D-0290 (PI: Kuncarayakti) and 079.D-0124 (PI: Kjaer).
The work by KM has been partly supported by Japan Society for the Promotion of Science (JSPS) KAKENHI Grant 26800100 (KM) and by World Premier International Research Center Initiative (WPI Initiative), MEXT, Japan. The authors acknowledge a support by JSPS Open Partnership Bilateral Joint Research Project between Japan and Chile (KM).
%This research is supported in part by the World Premier International Research Center Initiative (WPI Initiative), MEXT, Japan. 
KN is supported by the Grant-in-Aid for Scientific Research of the JSPS (23224004 and 26400222), Japan.

%The Acknowledgements section is not numbered. Here you can thank helpfulcolleagues, acknowledge funding agencies, telescopes and facilities used etc. Try to keep it short.

%%%%%%%%%%%%%%%%%%%%%%%%%%%%%%%%%%%%%%%%%%%%%%%%%%



%\bibliographystyle{mnras}


%%%%%%%%%%%%%%%%%%%%%%%%%%%%%%%%%%%%%%%%%%%%%%%%%%

%%%%%%%%%%%%%%%%% APPENDICES %%%%%%%%%%%%%%%%%%%%%

%\appendix

%%%%%%%%%%%%%%%%%%%%%%%%%%%%%%%%%%%%%%%%%%%%%%%%%%

% Don't change these lines
\bsp	% typesetting comment
\label{lastpage}
\end{document}